\documentclass[twocolumn]{aastex631}

\received{January 7, 2025}
\revised{January 30, 2025}
\accepted{January 31, 2025}

\shorttitle{Filling the gap: the missing eclipses of $\gamma$ Persei from 2005 and 2006}
\shortauthors{\'{A}d\'{a}m~\&~Moln\'{a}r}

\usepackage{natbib, twoopt}
\usepackage{textcomp, gensymb}
\usepackage{multirow}
\usepackage{longtable}
\usepackage{supertabular}
\usepackage{amsmath}
\usepackage{amssymb}
\usepackage[T1]{fontenc}
\usepackage{booktabs}
\usepackage{tabularx}
\usepackage{array}

\begin{document}

\title{Filling the gap: the missing eclipses of $\gamma$ Persei from 2005 and from 2006}

\correspondingauthor{Roz\'{a}lia~Z.~\'{A}d\'{a}m}
\email{adam.rozalia@csfk.org}

\author[0009-0005-5307-8885]{Roz\'{a}lia Z.~\'{A}d\'{a}m}
\affiliation{E\"{o}tv\"{o}s Lor\'{a}nd University, Department of Astronomy, P\'{a}zm\'{a}ny P\'{e}ter s\'{e}t\'{a}ny 1/A, H-1117 Budapest, Hungary}
\affiliation{Konkoly Observatory, Research Centre for Astronomy and Earth Sciences, HUN-REN, MTA Centre of Excellence, \\ Konkoly Thege Mikl\'os \'ut 15-17, H-1121 Budapest, Hungary}

\author[0000-0002-8159-1599]{L\'{a}szl\'{o}~Moln\'{a}r}
\affiliation{Konkoly Observatory, Research Centre for Astronomy and Earth Sciences, HUN-REN, MTA Centre of Excellence, \\ Konkoly Thege Mikl\'os \'ut 15-17, H-1121 Budapest, Hungary}
\affiliation{E\"otv\"os Lor\'and University, Institute of Physics and Astronomy, P\'azm\'any P\'eter s\'et\'any 1/A, H-1117 Budapest, Hungary}

\begin{abstract}
    $\gamma$ Persei is a long-period ($P \approx 14.6$ yr) eclipsing binary system. Its period makes it a difficult target to fully understand: so far, only two primary eclipses are known in the literature, from 1990 and from 2019, whereas the 2005 one was missed due to its closeness to the Sun at the time.
    We aimed to fill in this gap by processing the quasi-continuous photometry collected by the Solar Mass Ejection Imager (SMEI) between 2003 and 2011, which was ideally positioned to observe such a bright targets.
    In order to do that, we first determined a color-dependent conversion formula from the SMEI measurements into {\it Gaia} \textit{G} magnitudes. We applied various corrections to the photometry and provide the longest continuous light curve of $\gamma$ Persei.
    We successfully detected the 2005 primary eclipse of the system, with the yearly observations ending during the egress of the companion. We predicted the position of a possible secondary eclipse by forward modeling the binary system with PHOEBE, and successfully recovered the secondary eclipse in the 2006 SMEI observations. The existence of the secondary eclipse puts strong constraints on the orbital configuration, which will be an important constraint for future studies of the system. 
\end{abstract}

\keywords{Eclipsing binary stars (444) -- Photometry (1234) -- Computational methods (1965)}

\section{Introduction} \label{sec:intro}

The object $\gamma$ Persei is an eclipsing binary system with an almost 15 year-long orbital period. Its changing radial velocity and composite spectrum was discovered more than a century ago \citep{Campbell-1909}.  Since then, it has also been studied as a visual binary \citep{Hartkopf_2001} and a spectroscopic binary \citep{Pourbaix_2000, Pourbaix_2004, Picotti_2020, Diamant_2023}, respectively. Moreover, \cite{Pourbaix_1999} analyzed the radial velocity curve of the system, and published its orbital parameters.

It is a member of the $\zeta$~Aur class: double-lined binaries with a cold G-K giant primary and a hot secondary. They show `atmospheric eclipses', when the secondary goes behind the chromosphere of the giant, revealing the properties of it  \citep{Wright_1970,Ake_2015}.

Nevertheless, the extremely long period of the binary makes it a challenging target to study. So far, only two primary eclipses are known in the literature, from 1990 \citep{Griffin_1994} and 2019 \citep{Diamant_2023}.
Albeit being a bright system, easily observed by the naked eye, the closeness of the Sun made it impossible for usual photometric telescopes to measure its 2005 eclipse, as the star was only weeks away from conjunction.

The analysis of very long-period eclipsing binaries hinder us with major obstacles. The longer the period, the longer the semi-major axis, which also decreases the chance of the observation of such a companion. As such, they have very low eclipse probabilities and require extensive observations to sample a full orbit. There are only a handful of cases where eclipses were observed and those turned out to be caused by a circumstellar dust ring instead of a star. One famous example is the $\varepsilon$ Aurigae system with a $27.1$-year period, whose eclipsing body took almost two centuries to be understood \citep{Backman-1984,kloppenborg_2010}.  

In this paper, we aim to fill the gap in the eclipse coverage for $\gamma$ Persei by presenting newly processed observations. In Section~\ref{sec:data} first we introduce the instrument, then we describe our methods, while in Section~\ref{sec:results} we present our findings. Lastly, in Section~\ref{sec:discussion} we aim to summarize and highlight the importance of our results.

\section{Data Processing} \label{sec:data}

\subsection{The Solar Mass Ejection Imager instrument}
The Solar Mass Ejection Imager (SMEI, \citealt{Eyles_2003, Jackson_2004}) was a heliophysics instrument funded by NASA, the US Air Force, and the University of Birmingham. SMEI was launched on the US Department of Defense Space Test Program's \textit{Coriolis} satellite in January 2003 and was deactivated in September 2011, although the nominal length of the mission was only 3 years.

SMEI was designed as a proof-of-concept instrument for forecasting the arrival of coronal mass ejections and for exploring their role in the occurrence of geomagnetic storms \citep{Webb_2002}.
As emphasized in \cite{smei_review}, it was the first heliospheric imager, which is essentially a wide-angle version of a coronagraph.
The satellite had a Sun-synchronous polar orbit ($i\sim 98 \degree$) at an altitude of 840 km with a 102-minute period.

SMEI's field-of-view (FOV) was divided between three cameras, each covering $60\degree \times 3\degree$, which resulted in a combined instantaneous FOV of $\sim 160\degree \times 3\degree$.
Camera 1 observed between $\sim\!120\degree$--$180\degree$ (from the Sun), Camera 2 scanned the sky around approximately $70\degree$--$130\degree$, while Camera 3 observed around $18\degree$--$80\degree$. Therefore, each orbit covered almost all of the sky, except for a $\sim 18\degree$ radius exclusion region around the sunward pole of orbit and a smaller exclusion region in the anti-Sun direction.

The instrument suffered from various issues over its lifetime. The main instrumental factors were hot pixels and scattered light, while several environmental effects and astronomical phenomena (such as the Moon, zodiacal light, etc.) also contributed to the hardships the mission endured.
The performance of Camera 3 showed the highest decline, since the temperature of the camera was between $-15$ and $0\,\degree$C, instead of the planned $-30\,\degree$C.
Although not as heavily affected, the efficiency of Cameras 1 and 2 also decreased over the course of the observations, mostly as a result of dark charge noise from hot pixel accumulation.
Radiation damage to the CCDs also became an issue for these cameras later in the mission.

The observed passband was between $450$ and $950$ nm, where the quantum efficiency of the CCD was at least $10\%$. 
Exposure cadence was $4$ s.
Additionally to its main purpose, SMEI was capable of detecting stars down to 10th magnitude in a single sky mapping, too.
In order to analyze the coronal mass ejections, stars brighter than 6th magnitude were individually fitted and removed \citep{Hick-2007}.

We must note that SMEI was the first mission to provide multi-year, quasi-continuous, densely sampled photometry for the brightest stars, although at the cost of limited photometric precision.
Several astrophysical studies already took advantage of the instrument for various science cases, such as analyzing novae \citep{hounsell_2010}, the red supergiant Betelgeuse \citep{Joyce_2020, Jadlovsky_2023}, the Cepheid star, Polaris \citep{Bruntt-2008}, or other binary systems \citep{Bruntt_2006, Pribulla_2010}.

\subsection{SMEI intensity conversion to {\it Gaia} \textit{G}-band}
After accessing the raw SMEI photometry\footnote{The original SMEI webpage is not accessible anymore, but the complete dataset can be downloaded from this ftp: \url{ftp://cass185.ucsd.edu/smei_star}.}, we first converted the time units, to Barycentric Julian Dates (BJD). We then converted the SMEI intensities to {\it Gaia} \textit{G}-band magnitudes. We used all stars observed by SMEI, and crossmatched them with the {\it Gaia} DR3 catalog \citep{Gaia-EDR3-2021}. This resulted in $5257$ objects, for which we collected the median intensities measured by SMEI, the {\it Gaia} \textit{G} mean magnitudes and the \textit{BP--RP} colors.

\begin{figure*}
    \centering
    \includegraphics[width=\linewidth]{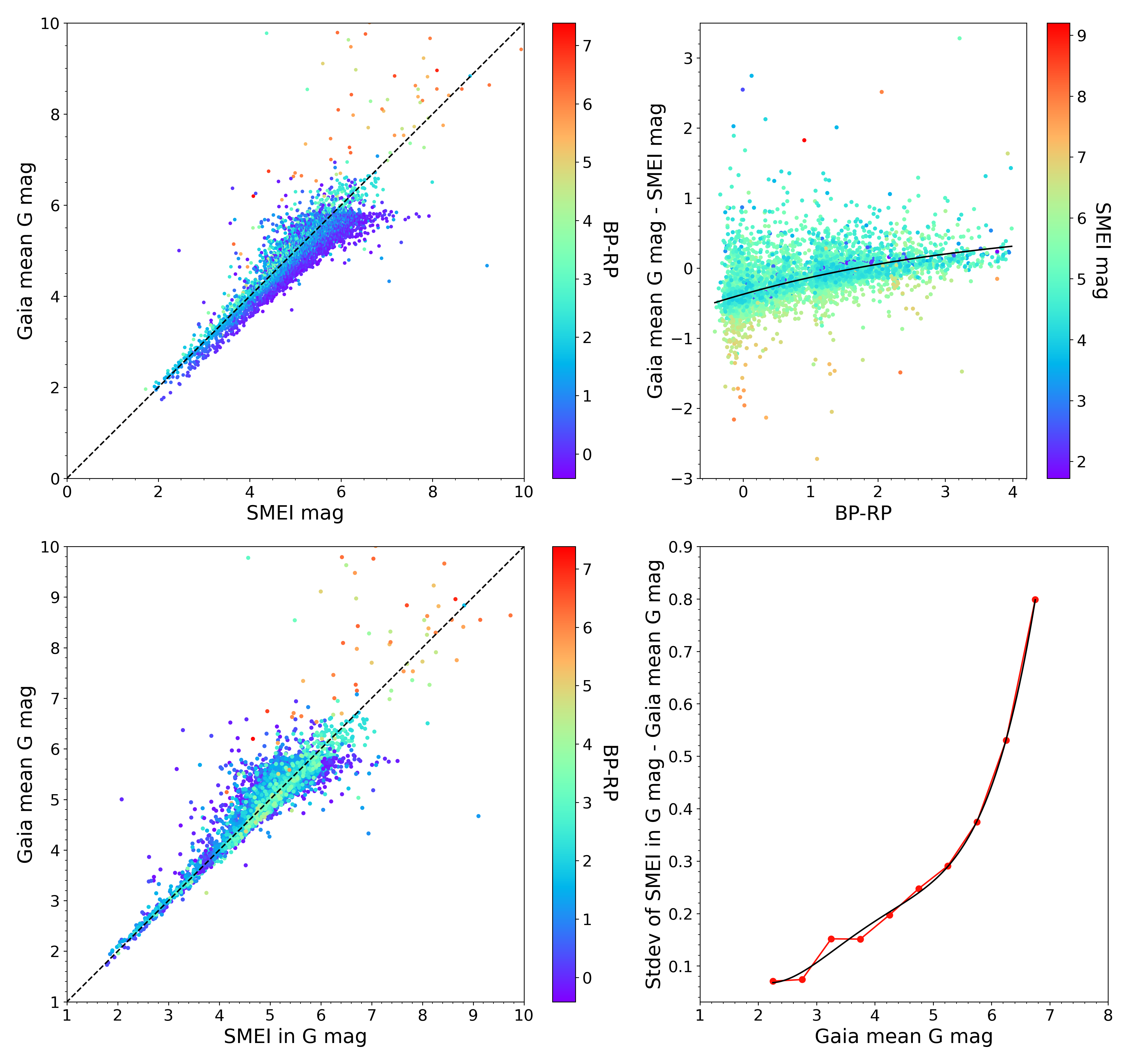}
    \caption{The upper left panel shows the {\it Gaia} DR3 mean \textit{G} magnitudes for every SMEI stellar target against the magnitudes calculated from the SMEI intensities and magnitude zero point ($m_0$) we determined. The black dashed line refers to the consistency between the different brightness values. A clear color dependence can be identified.
    Below, the same image is shown after the inclusion of color dependence into the conversion. Here, the data points spread nicely along the black dashed line.
    In the upper right, we fit the color dependence with a third-order polynomial.
    To represent the accuracy of our conversion, in the fourth image we show the standard deviation of the difference between the converted SMEI and the mean {\it Gaia} \textit{G} magnitudes with red. While the fourth-order polynomial fit of the trend is shown with black.
    }
    \label{fig:mag_conversion}
\end{figure*}

To convert the intensities into magnitudes, first we compared the {\it Gaia} \textit{G} mean magnitudes with the SMEI magnitudes (calculated with the Pogson formula). 
We determined the magnitude zero point to be $m_0 =8.97397$~mag. We then examined the color dependence of the sample in Fig.~\ref{fig:mag_conversion} upper left, where an apparent color bimodality can be seen.

To include this factor into the magnitude conversion, we expanded the Pogson formula with a color term:
\begin{equation}
    m_G = -2.5 \cdot \log I + m_0 + \Delta m_{\textup{BP--RP}}
    \label{eq:conversion}
\end{equation}
In Eq.~(\ref{eq:conversion}) $m_G$ is the {\it Gaia} \textit{G} magnitude, $I$ is the median intensity measured by SMEI and $m_0$ is the zero point.
$\Delta m_{\textup{BP--RP}}$ is the color dependence function, which is determined by fitting the SMEI magnitudes subtracted from {\it Gaia} \textit{G} magnitude values.
We note that the SMEI intensities are calculated as the product of the `standard star' flux (\texttt{Istd}) and the relative intensity of the target (\texttt{I/Istd}) found in the SMEI photometry tables (these are the $F^{\rm std}$ and $B$ quantities in Eq.~(5) in \citealt{Hick-2007}).

After the analysis, we concluded that a third-order polynomial provides the best fit for $\Delta m_{\textup{BP--RP}}$:
\begin{multline}
    \Delta m_{\textup{BP--RP}} = 0.0017 \cdot (\textup{BP--RP})^3 - 0.0318 \cdot (\textup{BP--RP})^2 \\ + 0.2715 \cdot (\textup{BP--RP}) - 0.3731
    \label{eq:deltam}
\end{multline}

The fitted polynomial is shown in the upper right plot of Fig.~\ref{fig:mag_conversion}. After correcting for the color dependence factor in Eq.~(\ref{eq:deltam}), we arrive at the lower left panel of Fig.~\ref{fig:mag_conversion}, where the \textit{BP--RP} bimodality issue has been resolved. To check the goodness of the conversion, we calculated the standard deviation of the difference between SMEI converted and {\it Gaia} \textit{G} mean magnitudes. In Fig.~\ref{fig:mag_conversion}, the trend of this difference is displayed with a fourth-order polynomial fit (see Eq.~(\ref{eq:stdev_fit})). This equation provides an estimate for the photometric accuracy of the conversion as a function of brightness:
\begin{multline}
    \sigma m_{\textup{Gaia}} = 0.0064 \cdot m_{\textup{Gaia}}^4 - 0.0976 \cdot m_{\textup{Gaia}}^3 \\ + 0.5480 \cdot m_{\textup{Gaia}}^2 - 1.2684 \cdot m_{\textup{Gaia}} + 1.0950
    \label{eq:stdev_fit}
\end{multline}

\subsection{SMEI light curve} \label{sec:lc_correction}
The data are stored in IDL tables, where the time in an unusual format (\texttt{year\_dayofyear\_hhmmss}). So we first we converted the time codes to BJD\_TDB values. We then made a criterion for \texttt{npsf}, the parameter describing the number of points used to determine the point spread function for each individual stellar measurement. As such, this parameter offers a way to quantify the goodness of the photometry. We discarded points where the relative value of \texttt{npsf} to the median was less than 0.9.
Additionally, we made initial cuts in intensity ($I<120$ and $I>400$) to eliminate further outliers.

The light curve showed various instrumental variations. These include a long-term decline in the average intensity, strong annual variations in each camera, offsets between the three cameras, as well as fast, short-timescale changes.

After plotting the observations of the three cameras, we decided to correct them separately, as they had shown distinct trends. We followed similar steps in all cases. First, we subtracted the long-term variation after fitting the data with a fourth-order polynomial in the case of Camera 1 and 2, and a third-order one for Camera 3. Next, we identified yearly trends by folding the light curves with a period of $365.24$ days. To subtract them, we first flattened the 1-year-long light curve sections individually with linear fits, and then made a template light curve by stacking these sections together (except for the one containing the primary eclipse). We subtracted the template from the light curve sections from each year, and further flattened the sections, if necessary. In the case of Camera 2, even the template could not account for a parabolic trend, which we removed with second-order polynomials. Finally, excluding the primary eclipse, we performed a $3\sigma$ clipping on the light curve. The final, processed light curve data is presented in Table~\ref{tab:lc}.

\renewcommand{\arraystretch}{1.3}
\begin{table}[htb!]
    \centering
    \caption{Sample table of the processed SMEI light curve of $\gamma$~Persei. The full table is available in machine-readable format in the electronic version.}
    \label{tab:lc}
    \begin{tabular}{l c c c}
        \hline \hline 
        BJD--2450000 (days) & $G$ mag & $\sigma G$ (mag) & SMEI mag \\ 
        \hline
        $2673.480057$ & $2.565$ & $0.078$ &  $-6.242$\\
        $2673.621156$ & $2.529$ & $0.076$ &  $-6.277$\\
        $2673.832865$ & $2.575$ & $0.078$ &  $-6.226$\\
        $2673.903380$ & $2.598$ & $0.079$ &  $-6.204$\\
        $2674.396998$ & $2.629$ & $0.081$ &  $-6.177$\\
        $2675.173371$ & $2.608$ & $0.080$ &  $-6.200$\\
        $2675.243909$ & $2.613$ & $0.080$ &  $-6.196$\\
        $2675.385079$ & $2.493$ & $0.074$ &  $-6.316$\\
        \multicolumn{4}{l}{...}\\
        \hline
    \end{tabular}
\end{table}

\subsection{PHOEBE modeling}
We forward modeled the system using the implemented {\it Gaia} \textit{G} passband in the Physics of Eclipsing Binaries (PHOEBE) 2.0 code \citep{Prsa_2016}.

For the modeling we used the published parameters of the system. The most recent work about $\gamma$ Persei was published by \citet{Diamant_2023}. From this paper we used the updated period of the system ($P$), the radii ($R$) and temperatures ($T$) of the components. After some initial tests, we chose a primary mass ($M$) that is smaller but provides a better fit, and it is within 2$\sigma$ of the value published by \citet{Diamant_2023}. 

\begin{figure}[htb!]
    \centering
    \includegraphics[width=\linewidth]{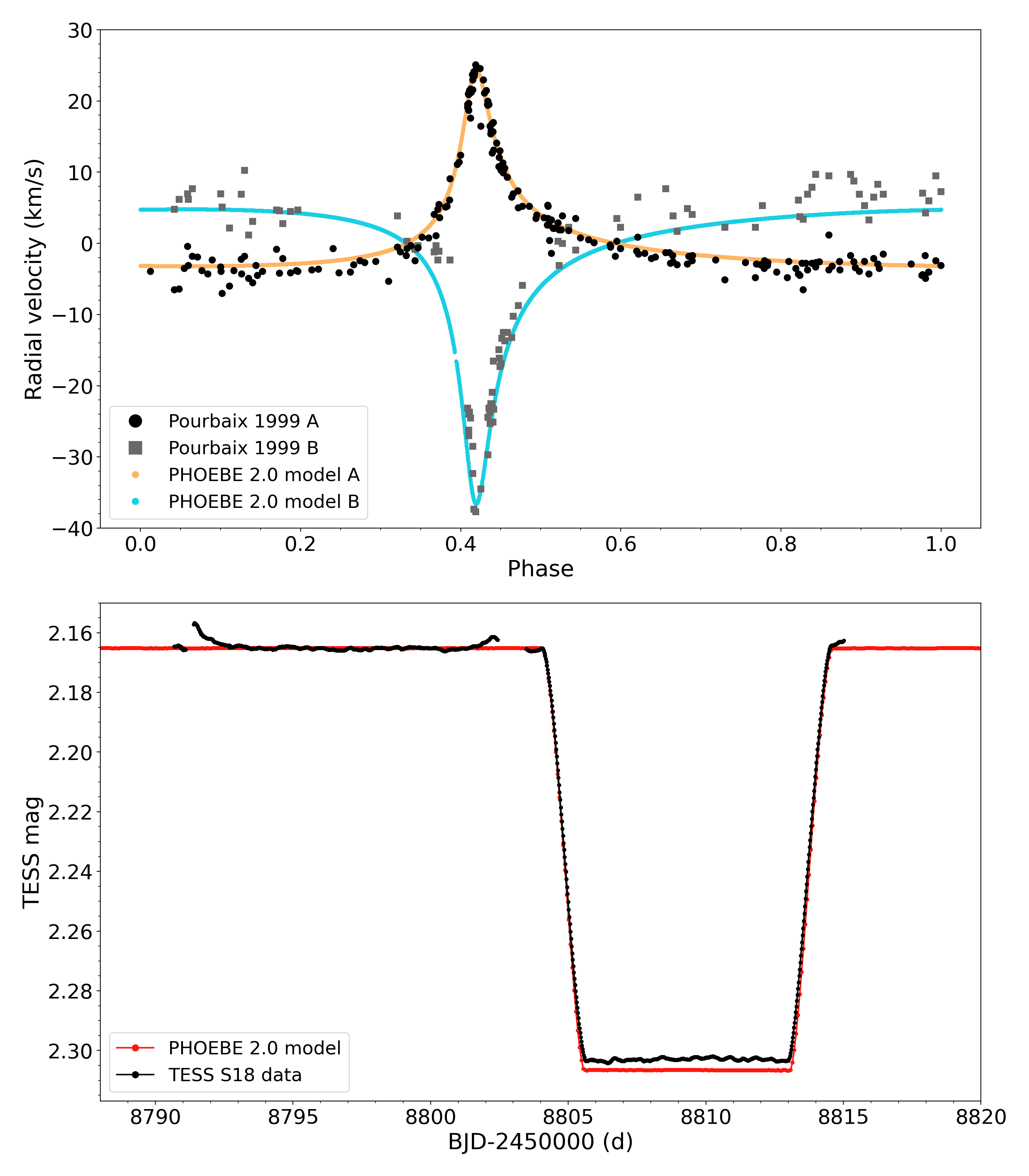}
    \caption{The upper panel shows historic RV measurements published in \cite{Pourbaix_1999} with the PHOEBE 2.0 model. The black dots denote the observed RV of the primary, while the gray squares the secondary component. The model is shown with colors, the light brown dots refer to the primary, while the blue denotes the secondary component.
    Below: TESS Sector 18 observations of the 2019 main eclipse (black) are presented with the PHOEBE 2.0 model (red) in the corresponding passband.}
    \label{fig:rv}
\end{figure}

To constrain the argument of periastron ($\omega$), we modeled the radial velocity (RV) curve of the system and compared it to historical RV data \citep{Pourbaix_1999}. Our conclusion was that the orbit needed half a rotation relative to \cite{Pourbaix_1999} in order to match the model with the data set, as can be seen in Fig.~\ref{fig:rv}. The obtained $\omega$ is within $3\sigma$ agreement with \cite{Popper_1987}.

We adopted the mass ratio ($q$) from \citet{Griffin_1994}.

Since the most precise observations of the system were collected by the Transiting Exoplanet Survey Satellite (TESS), we wanted to match those measurements with a PHOEBE model, as well. TESS observed the star only three times so far, but it serendipitously covered its 2019 main eclipse during Sector 18. We downloaded the Sector 18 2-minute cadence data \citep{t9-nmc8-f686} from the Mikulski Archive for Space Telescopes (MAST\footnote{\url{https://mast.stsci.edu/}}). The simple-aperture photometry (SAP) light curve of the star covers the complete eclipse, including the egress, although the ends of the TESS orbits are affected by scattered light. After comparing the model to the TESS photometry (see Fig.~\ref{fig:rv}), we chose to model a perfectly edge-on orbit ($i=90\degree$), which is within $1\sigma$ of the inclination of \cite{Pourbaix_1999}.
Our decision was also supported by the fact that this configuration automatically provides a secondary eclipse, which we intended to find.

We also reduced the eccentricity slightly, within $1\sigma$ as published by \cite{Pourbaix_1999}. Additionally, we adopted metallicity values from the literature. We set the [Fe/H] of the primary to $-0.19$ \citep{Soubiran_2022} and to $-0.2$ of the secondary \citep{Diamant_2023}.
The exact values of the aforementioned parameters are presented in Table~\ref{tab:model_parameters} with the corresponding references.

\renewcommand{\arraystretch}{1.3}
\begin{table}[htb!]
    \centering
    \caption{The astrophysical parameters used for the modeling with their references. The $1$s and $2$s in lower indices refer to primary and secondary component, respectively. The non-indexed values refer to the orbit.}
    \label{tab:model_parameters}
    \begin{tabular}{c | c | c}
        \hline \hline 
        \multirow{2}{*}{Parameters} & \multirow{2}{*}{Values} & \multirow{2}{*}{References} \\ 
        & & \\ \hline \hline
        $P$ & \multirow{2}{*}{$5329.08$} & \multirow{2}{*}{\cite{Diamant_2023}} \\
        \ [days] & & \\ \hline
         
        $R_1$ & $22.0$ & \multirow{2}{*}{\cite{Diamant_2023}} \\
        $R_2$ [$R_{\odot}$] & $3.7$ & \\  \hline
        
        $T_1$ & $5040$ & \multirow{2}{*}{\cite{Diamant_2023}} \\
        $T_2$ [K] & $8330$ & \\ \hline
        
        $M_1$ & \multirow{2}{*}{$3.3$} & $2\sigma$ of \\ \
        [$M_{\odot}$] & & \cite{Diamant_2023} \\ \hline
        \multirow{2}{*}{$q$} & \multirow{2}{*}{$0.666$} & \multirow{2}{*}{\cite{Griffin_1994}} \\
         & & \\ \hline
        \multirow{2}{*}{$e$} & \multirow{2}{*}{$0.781$} & $1\sigma$ of \\
        & & \cite{Pourbaix_1999} \\ \hline
        $i$ & \multirow{2}{*}{$90$} & $1\sigma$ of \\ \
        [$\degree$] & & \cite{Pourbaix_1999} \\ \hline
        $\omega$ & \multirow{2}{*}{$349.6$} & \cite{Pourbaix_2000} \\ \
        [$\degree$] & & (after a $180\degree$ rotation) \\ \hline
        \ [Fe/H]$_1$ & $-0.19$ & \cite{Diamant_2023} \\
        \ [Fe/H]$_2$ & $-0.2$ & \cite{Soubiran_2022} \\ \hline
    \end{tabular}
\end{table}

It must be noted that in order to fix the primary mass in our forward model, we flipped the constraint, so that we could solve for the semi-major axis instead of the mass of the primary.
Additionally, we increased the bolometric gravity brightening parameter of star A to \texttt{grav\_bol}$=0.9$, due to its high temperature. As suggested, we also set a higher value to reflection and heating parameter for the secondary, \texttt{irrad\_frac\_refl\_bol}$=1.0$.

\begin{figure*}
    \centering
    \includegraphics[width=\linewidth]{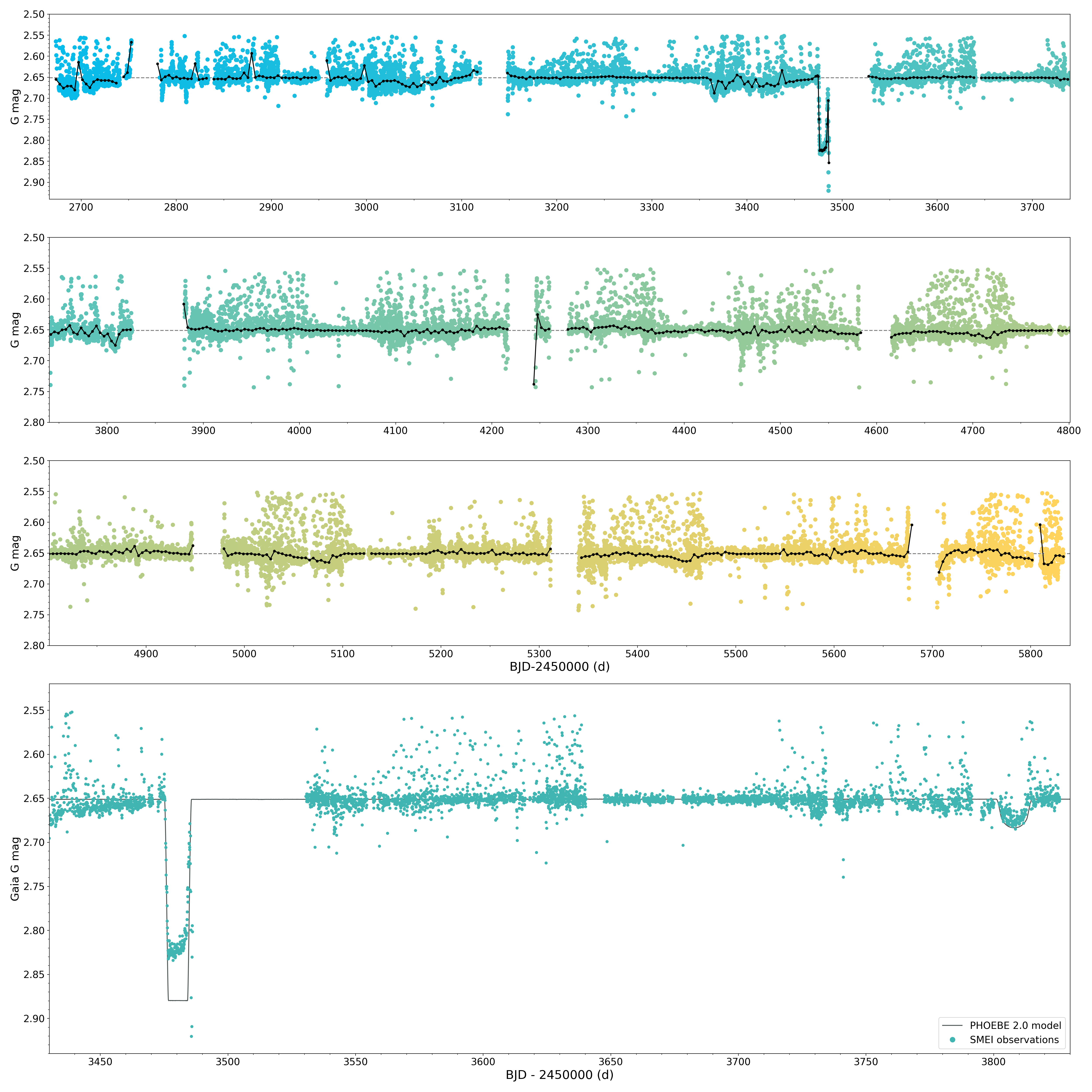}
    \caption{The upper image presents the longest continuous light curve of $\gamma$ Persei as observed by SMEI. The plot shows combined data from all the three cameras with a $3\sigma$ clip. The color is changing with time. Black dots refer to the binned light curve. The main eclipse is binned separately.
    Below a short segment of the same light curve; the primary and secondary eclipses of $\gamma$ Persei from 2005 and from 2006 (turquoise dots) plotted over the PHOEBE 2.0 model (dark gray line).}
    \label{fig:smei_lc_long}
\end{figure*}

\section{Results} \label{sec:results}

After following the steps described in Section~\ref{sec:lc_correction}, we were presented with the longest continuous light curve of $\gamma$ Per (see Fig.~\ref{fig:smei_lc_long} upper image), where the primary eclipse is clearly visible between BJD$2453475$ and BJD$2453485$. In Fig.~\ref{fig:smei_lc_long} we combined all observations from the three cameras after doing the corrections separately. In the upper panel, the black dots refer to the binned light curve, where the main eclipse was binned separately, with more frequent bins. The eclipse happened right at the end of the observing window of Camera 2. The second part of the eclipse and the egress was clearly affected by the decreasing separation from the Sun. Nevertheless, the event was recorded by SMEI right up until the fourth contact and the end of the eclipse.

After modeling the system with PHOEBE in the corresponding {\it Gaia} \textit{G} passband, we matched the model with observations to search for signs of a secondary eclipse. A small dip is indeed visible after BJD$2453800$, this feature does not repeat in other years. This brightness minimum also aligns perfectly with the model predicted secondary eclipse, making it easily identifiable in the right end of the lower panel of  Fig.~\ref{fig:smei_lc_long}. We measure the depths of the eclipses to be $0.179$~mag, measured at the beginning of the main eclipse to avoid systematics in the latter stages, and $0.027$~mag in the middle of the secondary eclipse.

The depths of the eclipses are smaller than what the model predicts. This can be caused by a combination of different effects. The uncertainty coming from the {\it Gaia} conversion, which at the brightness of $\gamma$~Per is $\sigma G = 0.081$~mag, exceeds the difference between the model and the observations in itself. The light curve could also be contaminated by nearby fainter sources, but there are no stars brighter than 8.0 mag in \textit{G}-band within $0\overset{^\circ}{.}75$ (the mission's limit for overlapping stars), so this effect is likely negligible. Finally, the eclipse depth might be offset due to limitations of the binary model we use to predict the position of the secondary eclipse. We also note that \citet{Joyce_2020} also found the variation amplitude of $\alpha$~Ori to be smaller in the SMEI photometry than in the ground-based $V$-band observations.

The time difference between the eclipses is $327.405$ days, which corresponds to an orbital phase difference of $0.061$. This aligns with our expectations, since this system has been identified as a highly eccentric binary with $e=0.785$ \citep{Pourbaix_1999}. From this we can calculate the times of later secondary eclipses. The following one happened in October 2020, with a midpoint at BJD$2459136.72$, but we are not aware of any observations. 

We predict the next primary and secondary eclipse mid-times to happen about a decade from now, on June 24, 2034 and May 18, 2035, respectively.

\section{Discussion} \label{sec:discussion}
The $\gamma$ Persei binary system has been known for a long time. It has been studied greatly due to it being multifaceted, meaning it is a visual, spectroscopic and an eclipsing binary, all at once. However, the extremely long period of the system hindered the determination of its precise astrophysical parameters. \cite{Griffin_1994} presented the primary eclipse from 1990, while \cite{Diamant_2023} analyzed the primary eclipse from 2019.

Here, we present the missing primary eclipse from 2005 and the first ever recorded secondary eclipse from 2006 as observed by the SMEI instrument. First, we determined a conversion between the SMEI intensities and the {\it Gaia} \textit{G}-band magnitudes. Then we processed the observations, and corrected the light curves for a number of instrumental systematics. Finally, we transformed the corrected intensities into the {\it Gaia} \textit{G}-band. The final light curve clearly shows the missing 2005 eclipse, with the observations ending during the egress of the secondary. 

We made forward models with the PHOEBE 2.0 python code, using the TESS eclipse observations and historical RV data as references. After we fixed the model parameters, we matched the models with the SMEI observations. With the help of this model, we successfully identified the first secondary eclipse observed in this system (concerning the literature). Unfortunately, the next secondary eclipse in 2020 (to our knowledge) went unobserved. 

Our model solution includes an $i=90\degree$ edge-on system to find the secondary eclipse. \cite{Griffin_1994} shows that the main eclipse is not central, but controlling for this through the orbital inclination is difficult at such a wide orbit. Future modeling of the system, however, should combine these two criteria (non-zero impact parameter and a secondary eclipse) into a single, strongly constrained solution. 

With the addition of the observation of the secondary eclipse, now there is a greater chance of determining the most accurate parameters of $\gamma$ Per. This is especially relevant from a stellar evolutionary standpoint, since \citet{Pourbaix_1999} hypothesized that the main component might be a merger product. In a future work, we aim to achieve this task and present a detailed analysis of the system, using further TESS photometry and high-resolution RV measurements.

\section*{Acknowledgments}
This project was supported by the KKP-137523 `SeismoLab' \'Elvonal grant of the Hungarian Research, Development and Innovation Office (NKFIH). This paper includes data collected with the TESS mission, obtained from the MAST data archive at the Space Telescope Science Institute (STScI).
Funding for the TESS mission is provided by the NASA Science Mission Directorate. STScI is operated by the Association of Universities for Research in Astronomy, Inc., under NASA contract NAS 5–26555.
This work has made use of data from the European Space Agency (ESA) mission {\it Gaia} (\url{https://www.cosmos.esa.int/gaia}), processed by the {\it Gaia} Data Processing and Analysis Consortium (DPAC, \url{https://www.cosmos.esa.int/web/gaia/dpac/consortium}).
Funding for the DPAC has been provided by national institutions, in particular the institutions participating in the {\it Gaia} Multilateral Agreement. This research has made use of NASA's Astrophysics Data System.

\software{Astropy \citep{astropy:2013, astropy:2018, astropy:2022}, PHOEBE 2.0+ \citep{Prsa_2016, phoebe_2018, phoebe_2020_1, phoebe_2020_2}}

\bibliographystyle{aasjournal}
\bibliography{references}

\end{document}